# Seeing Around Obstacles with Terahertz Waves

Yiran Cui[1] and Georgios C. Trichopoulos[1]

[1]School of Electrical, Computer and Energy Engineering, Arizona State University, Tempe, USA

Traditional imaging systems, such as the eye or cameras, image scenes that lie in the direct line-of-sight (LoS). Most objects are opaque in the optical and infrared regimes and can limit dramatically the field of view (FoV). Current approaches to see around occlusions exploit the multireflection propagation of signals from neighboring surfaces either in the microwave or the optical bands. Using lower frequency signals anatomical information is limited and images suffer from clutter while optical systems encounter diffuse scattering from most surfaces and suffer from path loss, thus limiting the imaging distance. In this work, we show that terahertz (THz) waves can be used to extend visibility to non-line-of-sight (NLoS) while combining the advantages of both spectra. The material properties and roughness of most building surfaces allow for a unique combination of both diffuse and strong specular scattering. As a result, most building surfaces behave as lossy mirrors that enable propagation paths between a THz camera and the NLoS scenes. We propose a mirror folding algorithm that tracks the multireflection propagation of THz waves to 1) correct the image from cluttering and 2) see around occlusions without a priori knowledge of the scene geometry and material properties. To validate the feasibility of the proposed NLoS imaging approach, we carried out a numerical analysis and developed two THz imaging systems to demonstrate real-world NLoS imaging experiments in sub-THz bands (270-300 GHz). The results show the capability of THz radar imaging systems to recover both the geometry and pose of LoS and NLoS objects with centimeter-scale resolution in various multipath propagation scenarios. THz NLoS imaging can operate in low visibility conditions (e.g., night, strong ambient light, smoke) and uses computationally inexpensive image reconstruction algorithms.

Due to the opaqueness of most materials in the visible and infrared portion of the electromagnetic (EM) spectrum, human vision, as well as regular optical cameras, are capable of seeing objects that are in the direct line-of-sight (LoS). However, when opaque surfaces are present in the scene, occluded objects are not visible, thus limiting the available field of view (FoV). On the other hand, lower frequency EM waves can penetrate most non-metallic or low loss materials and hence can reveal the anatomy and pose of non-line-of-sight (NLoS) objects and scenes. For example, through-the-wall-radars (TTWR) can detect the presence of humans or discern large objects behind concrete walls or other materials that are opaque in the visible spectrum [1-3]. Additionally, microwave frequencies have been used to detect NLoS objects through multireflection of the signals on surrounding opaque walls [4-6]. Nevertheless, the large wavelength of the low frequency signals and the limited bandwidth (typically <6 GHz) merely enable detection and tracking rather than clear images with detailed information. Additionally, the transparency of most materials, the small roughness of most surfaces with respect to the wavelength, and edge diffraction result in strong multipath propagation that clutters the reconstructed images. Therefore, microwave NLoS systems can be very effective for sensing hidden objects but offer limited imaging capabilities.

At infrared and optical frequencies, most materials are opaque and surfaces scatter most of the incident waves. As such, an alternative approach of NLoS imaging has been developed by exploiting the multipath propagation of lights, as first introduced in [7] where the authors demonstrate imaging of an occluded object by light scattering via a white, near-Lambertian surface. The method uses a laser for illumination and an ultra-fast infrared camera to record the time-of-flight of photons and reconstructs the image using a filtered backprojection-based algorithm. Gariepy et al showed a simpler approach to detect NLoS objects [8] and O'Toole et al. demonstrated a confocal NLoS imaging approach that can extend the visible distance; however, it mainly relies on the use of retroreflective surfaces [9]. Liu et al. developed a virtual camera concept to reduce image reconstruction complexity and account for scenes with multiple diffuse reflections [10]. Despite the recent advances [11-13], optical/infrared NLoS imaging needs to contend with several challenges: 1) limited by low visibility conditions (e.g., dust, smoke, fog), low reflectance surfaces (e.g., dark colors), and ambient lights (e.g., outdoors during daytime); 2) it usually requires complex and large systems to acquire backscattered signals; 3) the image reconstruction algorithms are computationally intensive; and 4) the diffuse scattering on the Lambertian surfaces suffers from a large loss and hence limits the imaging range [14].

As an emerging technology, the terahertz (THz) spectrum (0.3-3 THz) offers unique wave propagation phenomena that can alleviate many shortcomings of current NLoS imaging approaches. As illustrated in Fig. 1, at lower frequencies (<70 GHz) most building surfaces are smooth compared to the wavelength with negligible diffuse scattering. On the other hand, as previous works have shown, many surfaces can be characterized as near-Lambertian such that diffuse scattering dominates the wave reflection in the optical range. However, in the THz regime, the roughness of most building surfaces is comparable to the wavelength so that wave reflection is characterized by strong specular components as well as diffuse scattering, as it has been shown in [15]. Thereby, the *specular reflection* can be used to establish a strong signal propagation path between the imaging aperture and the NLoS scene, whereas the *diffuse scattering* can help reconstruct the shape of the LoS objects. So together they can be exploited for robust THz NLoS imaging without a priori knowledge of the scene geometry [16]. As shown below in this work, these unique scattering phenomena allow computationally efficient THz NLoS imaging with a centimeter-scale resolution and possibly long-distance imaging due to the strong specular reflections.

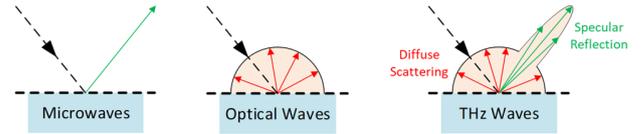

Figure 1. Wave scattering mechanisms across the EM spectrum.

The advances in semiconductor technology and nanofabrication techniques have enabled a plethora of THz active imaging systems that can generate high-resolution 2D and 3D images. Such systems include real-time THz cameras [17], stand-off raster scanners [18], as well as multiple-input-multiple-output (MIMO) imaging radars [19]. These systems can use moderate sized apertures to generate centimeter-scale resolution images at a >20 meters distance. So far, THz imaging research has focused on astronomy, imaging in low visibility conditions, spectroscopy, and seeing through light clothing for contrabands and security. Extending the vision to the NLoS and seeing around obstacles with THz waves can benefit multiple applications such as first response and rescue missions, detection for hidden traffic/pedestrians, and autonomous navigation. Besides, NLoS THz imaging can take advantage of the expected co-existence of radar and communication systems to provide high resolution simultaneous localization and mapping (SLAM) for future wireless communication systems [20].

In this paper, we present an image reconstruction method that can be applied to multireflection NLoS imaging (*mirror folding*). The method is evaluated first using a set of computationally generated 2D radar images. Then, we present measurement results of several experiments in real-life scenarios to further demonstrate the feasibility of THz NLoS imaging. We show that both the LoS and NLoS scenes can be successfully recovered with a centimeter-scale resolution.

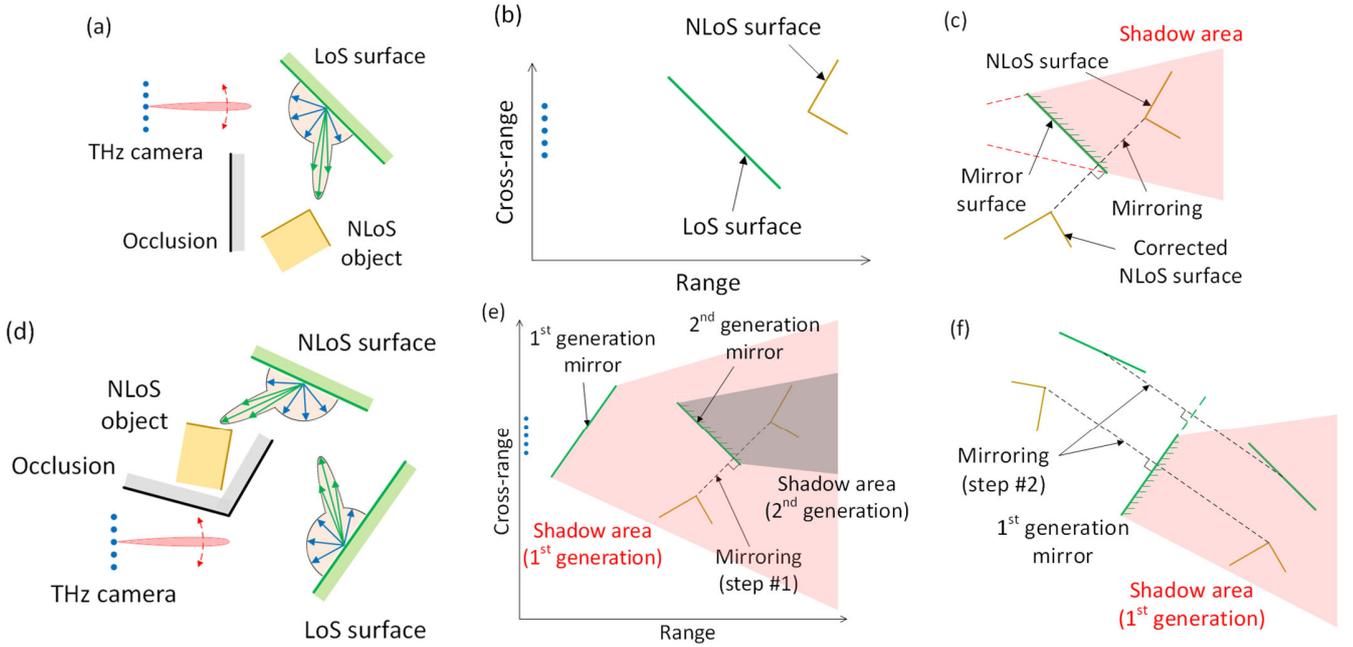

Figure 2 **Mirror folding for reconstruction of NLoS images.** (a) Illustration of a THz NLoS imaging scenario. Surrounding surfaces are used as mirrors to image the hidden scene behind nontransparent occlusions. (b) Due to the strong specular reflection from the LoS surface, the NLoS object appears as a "ghost" with a wrong pose in the initial image. (c) Assuming the LoS surface behaves as a mirror, the NLoS pose is corrected using *mirror folding*. (d) When multiple reflections between the LoS and NLoS objects are involved, more than one "ghost" objects appear in the initial image. (e-f) In this case, *mirror folding* is performed multiple times successively to properly determine the correct geometries.

## Results

Without loss of generality, we consider active THz imaging systems that generate narrow beams to scan a wide field of view (FoV) and use a finite bandwidth signal to acquire the range information of the backscattered signals (e.g., pulsed or continuous wave systems). The proposed method can be applied to THz images acquired by alternative systems, including MIMO-based imagers [19], programmable metasurfaces [21], or compressive sensing [22].

**THz radar imaging and mirror folding.** Let us consider an NLoS imaging scenario, as depicted in Fig. 2a. To image the invisible scene behind an opaque occlusion, an imaging aperture (THz camera) forms narrow transmitting and receiving beams that illuminate and record the backscattered signals from a LoS surface, respectively. Assuming the LoS surface is a common indoor building material, wave scattering features both strong specular reflection (green arrows) and detectable diffuse scattering that is almost omnidirectional (blue arrows). The latter enables imaging flat or lightly curved surfaces that are not normal to the imaging system's beam direction. On the other hand, the specular reflection allows a relatively low-loss roundtrip path between the aperture and the NLoS area and hence enables seeing around occlusions. In other words, the surrounding surfaces can act as *lossy mirrors* in the THz band and allow a relatively simple image reconstruction algorithm of NLoS scenes.

We assume an active imaging system where a narrow beam is scanned in both azimuth and elevation angles to obtain the angle-of-arrival (AoA) of the backscattered signals. In the meantime, at each scanning angle, the aperture measures the time-of-flight (ToF) information of the received signals. By combining them, a 3D image (range and cross-range) of the imaging scene is formed where the value of every voxel corresponds to the intensity of the backscattered signal. We note that 1) based on the Rayleigh criterion, the angular resolution is $\theta = 1.22\lambda/D$, where $\lambda$ and $D$ are the wavelength and aperture diameter, respectively; and 2) according to the sampling theorem, the range resolution is $c_0/(2B)$, where $c_0$ and $B$ are the speed of light in free space and measurement frequency bandwidth, respectively.

However, due to the nature of THz signal propagation, the 3D image may comprise backscattered signals originating from both single reflections (i.e., camera → LoS object → camera) and multiple reflections (e.g., camera → LoS object → NLoS object → LoS object → camera), as illustrated in Fig. 2a. As a result, although LoS objects can be imaged properly, due to longer propagation paths, signals from NLoS scenes arrive later and appear behind LoS objects with the wrong *pose* (location and orientation) in the initial image, as depicted in Fig. 2b. This is also known in radars as cluttering and results in false representation of the scene's geometry.

To correct the NLoS geometries and achieve an accurate image, we assume that, at THz frequencies, 1) all surfaces are opaque and 2) the specular reflection dominates the surface wave scattering. With these assumptions, most building surfaces can be treated as lossy mirrors with the amount of loss depending on the surface roughness and the material properties. This way, the whole area behind a LoS surface (determined by the relative position to the THz camera) is characterized as a "shadow area" (as illustrated in Fig. 2c). Correspondingly, the presence of any object inside this shadow area, classified as "ghost objects", is the result of a secondary reflection.

To account for this propagation, we mirror the objects inside the shadow area around the plane defined by the LoS surface (i.e., mirror) and obtain an image that depicts the NLoS objects with the correct geometries and poses, as shown in Fig. 2c. We term this operation as *mirror folding* since it reverses the process of mirror unfolding [23]. This method is also computationally efficient as it acts directly on the voxels of the image, thus adding a small overhead to the overall imaging process. On the contrary, in optical NLoS imaging approaches, the large amount of data being processed, and the complex algorithms lead to computationally intensive image reconstructions. For example, the back-projection method can take hours to produce images [7][24].

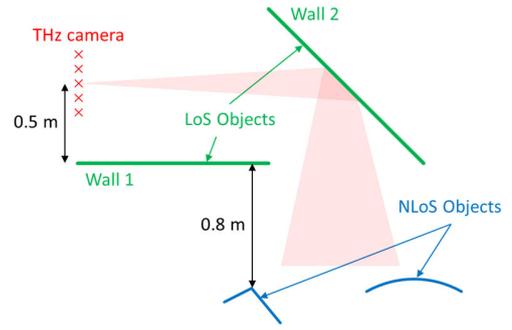

Figure 3. 2D geometry of a NLoS imaging scenario. The complex geometry results in multireflection propagation and image cluttering.

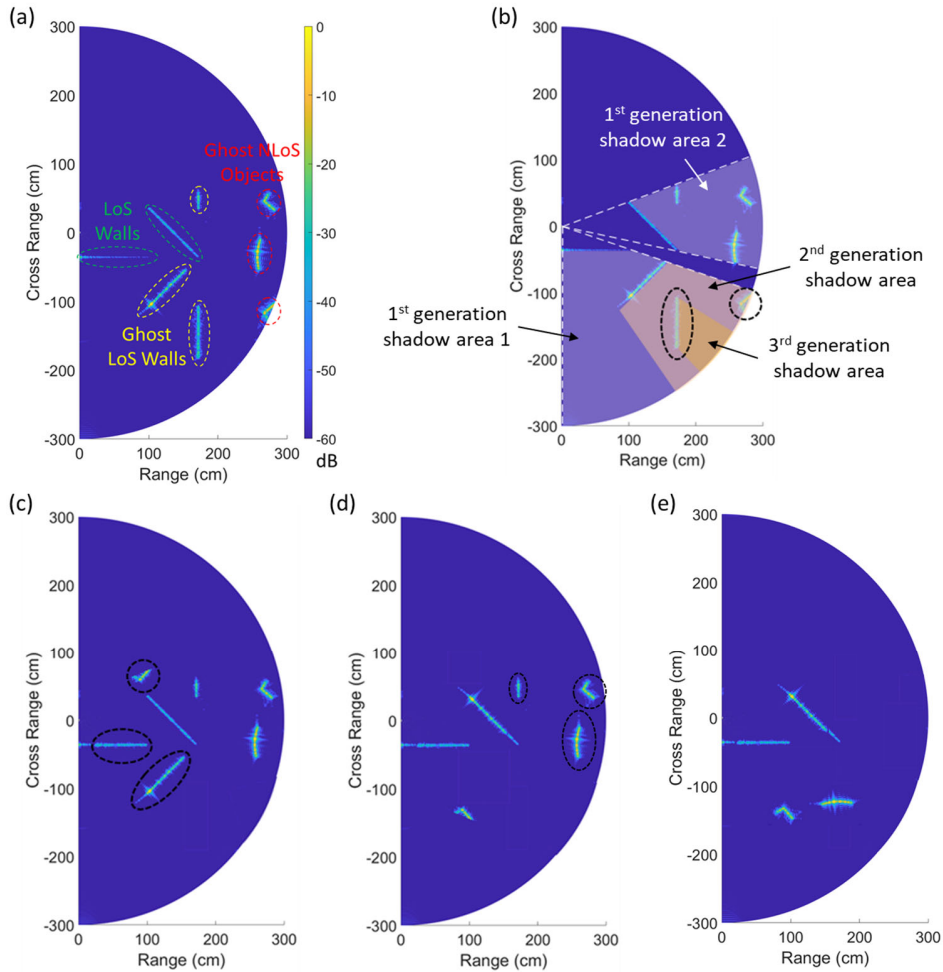

Figure 4. **Simulation of a 2D THz NLoS imaging scenario** (shown in Fig. 3). (a) The acquired (initial) image depicts several LoS walls as well as multiple "ghost" objects due to multireflection propagation of the THz waves. (b) The pixels are hierarchically classified into mirrors and objects, depending on the number of signal bounces needed to be imaged. LoS mirrors constitute the parents of the pixels located behind them, such that the first NLoS mirror is a $1^{st}$ generation mirror. Generations increase with the number of bounces. To properly reconstruct the NLoS scenes, *mirror folding* is applied multiple times beginning from the last generation toward the parent (LoS) mirrors. We notice that the $3^{rd}$ generation shadow area does not contain any objects. As such, the $2^{nd}$ generation pixels inside the purple marked area are mirrored around the respective mirror. (c) Then, these mirrored pixels (marked) "age" and move to the previous generation. As a result, they are mirrored again but around the parent mirror. (d) Similarly, the $1^{st}$ generation pixels (marked) are mirrored around the $2^{nd}$ parent (LoS) mirror to finally form (e) the corrected image that recovers the geometries of the original geometry.

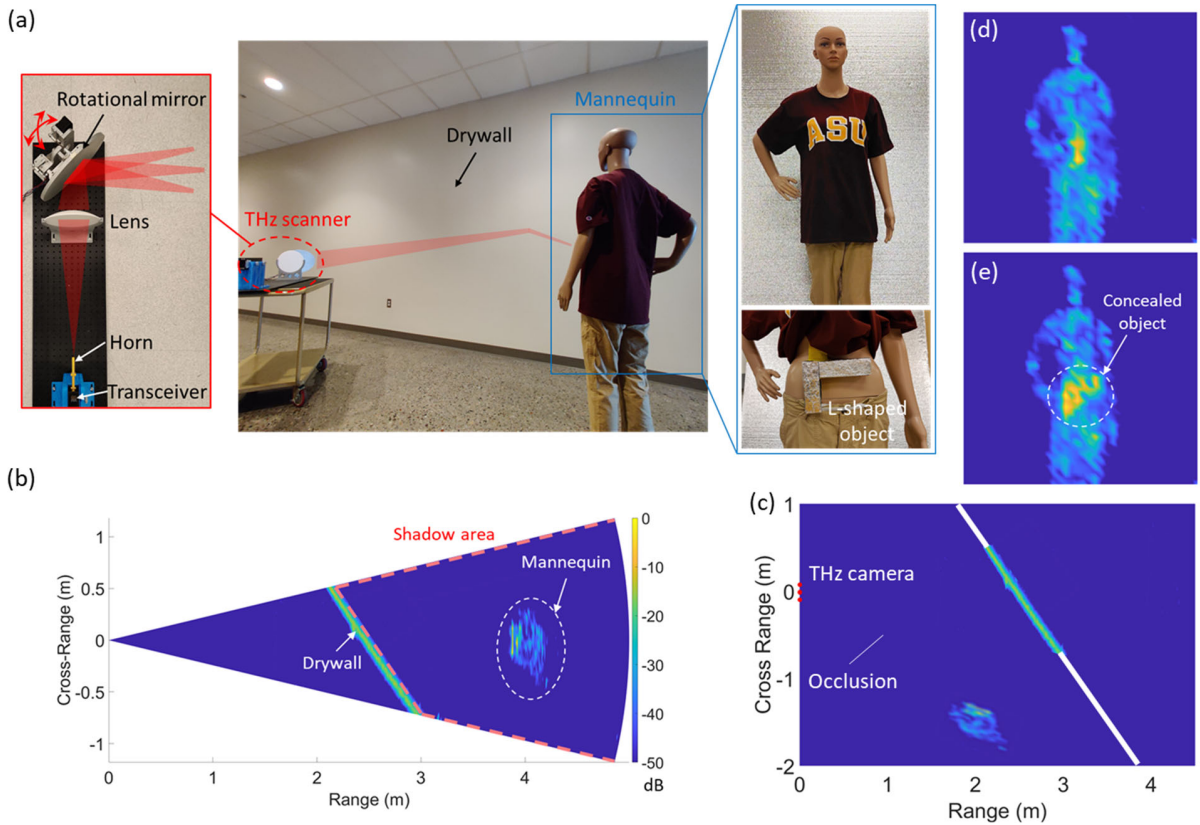

Figure 5.  **Concealed object detection in THz NLoS imaging.** (a) Measurement setup: a drywall in the LoS is used to image a hidden mannequin (occlusion not included). The mannequin carries a concealed L-shaped metallic object under the shirt. (b) A horizontal cross-section of the reconstructed image. The mannequin appears to be behind the drywall. (c) Using *mirror folding*, a correct image that agrees with the actual topology is obtained. (d-e) Vertical cross-sections before and after removing the concealed object showing the posture of the mannequin.

Nevertheless, high permittivity and/or high conductivity surfaces can allow multiple wave reflections within the scene of interest (SoI). Such propagation, as illustrated in the example of Fig. 2d, provides an opportunity to extend NLoS imaging to larger scenes but also can result in additional cluttering of the image. In this case, we need to perform *mirror folding* multiple times to obtain the correct image. As depicted in Fig. 2e and f, both the LoS and NLoS surfaces behave as mirrors in the initial image. To facilitate the multistep reconstruction process, we classify the hidden mirrors and objects depending on the number of bounces needed to image them. As such, the hidden mirror in this example is classified as a *child mirror* to the LoS wall and the object as children to the hidden mirror. Then, mirror folding is first applied to the children objects of the "youngest" mirror and then repeated for the LoS mirror. This hierarchical reconstruction process can be generalized for arbitrary number of bounces and is discussed in more detail in the methods section.

**THz NLoS imaging simulation.** Simulating an NLoS imaging scenario with a full-wave model is computationally expensive in the THz bands due to electrically large dimensions. For example, one meter corresponds to 1,000 λ at 300 GHz (λ is the wavelength), and we would need models that have a total surface area of multiple square meters. Thus, computing the fields propagation in most THz NLoS scenarios becomes prohibitively demanding in terms of CPU and RAM resources. An alternative approach is to reduce 3D geometries that are uniform in one dimension (e.g., height) to a 2D space [25]. For example, a circle in 2D corresponds to an infinitely long cylinder in 3D.

We assume a 2D imaging problem with both LoS and NLoS scenes and a THz camera, as depicted in Fig. 3. In this example, the scene comprises two LoS walls (both are 1 meter long) and two NLoS targets (a corner with a total length of 40 cm and a 50 cm-long arc). Wall 1 blocks the direct LoS between the THz camera and the NLoS targets, thus wall 2 is used to image the hidden scene.

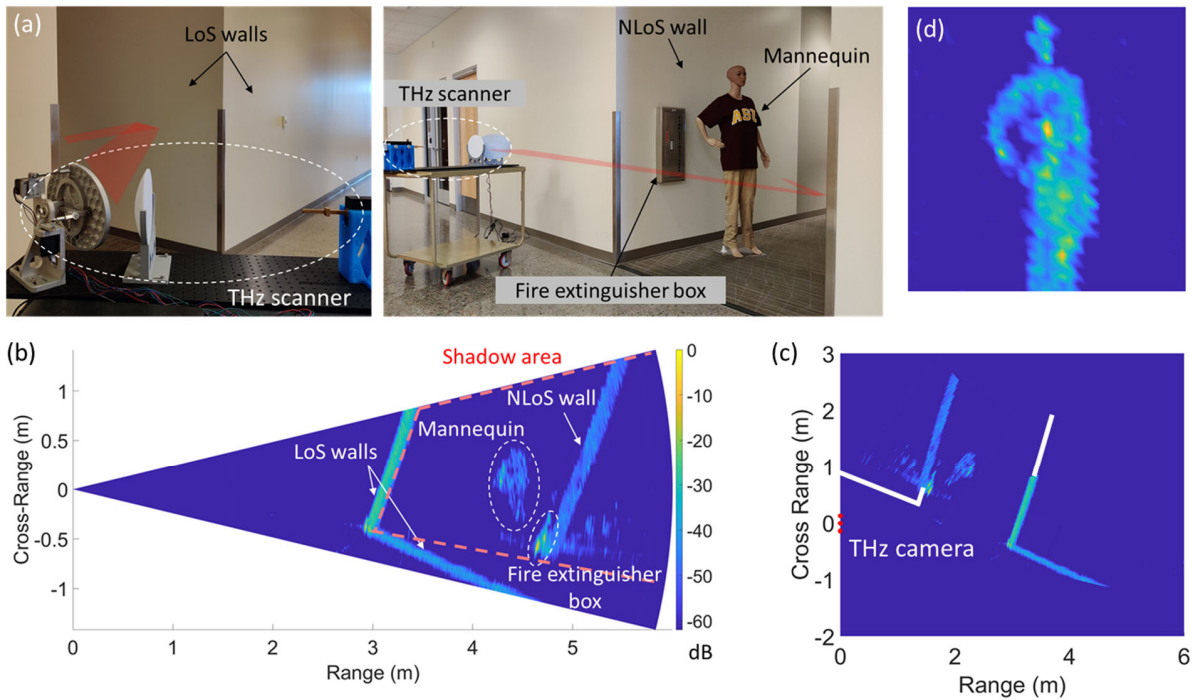

Figure 6. **Seeing around corners:** (a) A view of the scene under test from an optical camera at the THz imaging system's vantage point: Only the LoS walls are visible whereas a mannequin and other parts of the hallway are in the NLoS. (b) A horizontal cross-section of the reconstructed image. The NLoS objects appear to be behind one of the LoS walls. (c) Using *mirror folding*, a correct image that agrees with the actual topology is obtained. (d) A vertical cross-section clearly shows the mannequins geometry and pose.

Fig. 4a shows the obtained initial image using a frequency band of 290-300 GHz. We notice that 1) two LoS walls are correctly reconstructed (marked in green); 2) both LoS walls generate "ghosts" (marked in yellow) due to the multiple reflections between themselves; and 3) NLoS objects also create "ghosts" (marked in red), the two at the top are resulted from the right-handed wall whereas the one at the bottom is created due to the multiple reflections between the walls. Then, following the image correction process discussed above, we mirror the shadow areas behind the LoS walls (shown in Fig. 4b) to determine the accurate positions and orientations of the NLoS objects. As such, Fig. 4c-e depict the image correction steps using *mirror folding*. The corrected image, as depicted in Fig. 4e, agrees with the actual topology of the simulation model (see Fig. 3).

**THz NLoS imaging experiment.** To demonstrate THz NLoS imaging, we carried out experiments for various propagation scenarios. Two imaging systems, based on raster scanning and synthetic apertures, are used to investigate 3D and 2D NLoS imaging, respectively.

First, we implement a THz NLoS experiment using a 3D THz radar imaging system, as depicted in Fig. 5a. A drywall is used to image a life-sized mannequin placed in the NLoS. The imaging system uses optics (i.e., a lens and a rotational mirror) to generate a narrow THz beam that scans in both azimuth and elevation directions (approximately ±15° range). Thus, the scanner stares at the drywall at an oblique angle such that the signals are reflected toward the NLoS scene. As such, using a measurement frequency band of 270-300 GHz, we obtain an initial image as depicted in Fig. 5b (horizontal cross-section). We notice that the drywall is at the expected location and orientation. It appears to be thicker compared with the simulation results because the simulation does not account for the effect of volume scattering that occurs within a shallow depth inside the drywall material [26]. In addition, the shape of the mannequin is clearly reconstructed behind the drywall due to the strong specular reflection, however with a wrong pose. The correct image obtained using the *mirror folding* method is presented in Fig. 5c, we notice that it agrees with the actual measurement setup topology. Moreover, a vertical cross-section of the mannequin is shown in Fig. 5d, which clearly shows its outline and posture.

To highlight the relatively low-loss propagation under specular reflection, we present a scenario where the mannequin carries an L-shaped object (covered with aluminum foil to increase the reflectivity) under clothing. As depicted in the vertical cross-section of the mannequin (Fig. 5e), we can identify the concealed (i.e., covered by the mannequin's clothing) object appearing as a brighter area due to a stronger backscattering. Therefore, THz NLoS imaging has the potential to reveal concealed objects under clothing, which is beneficial to security applications.

Next, we deploy this 3D THz radar system to a more complicated NLoS imaging scenario, as shown in Fig. 6a.

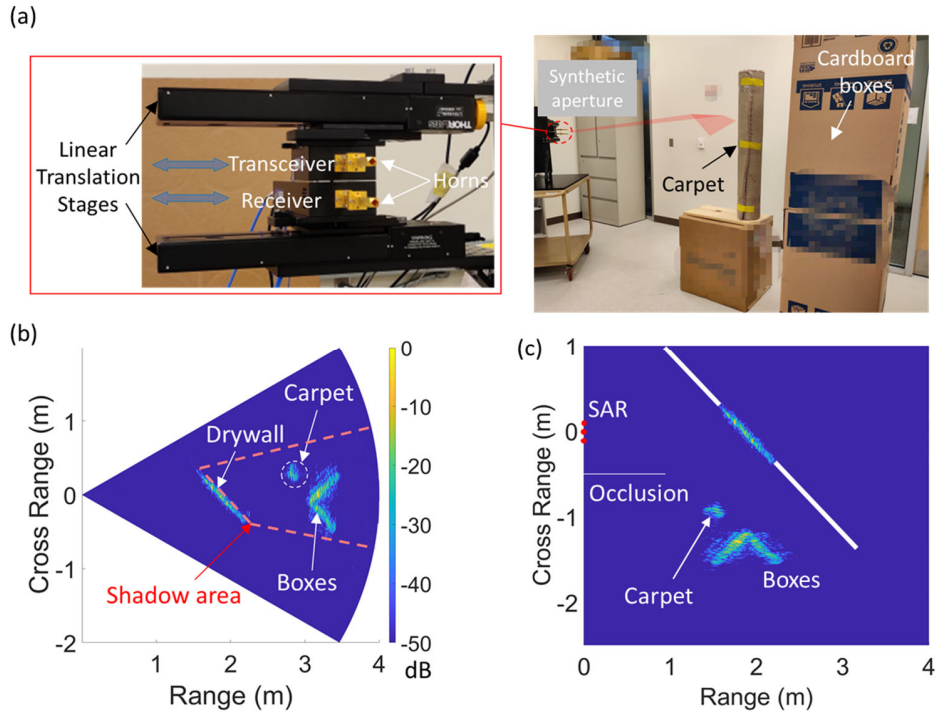

Figure 7. **THz NLoS imaging experiment using synthetic apertures.** (a) A picture of the measurement setup where a drywall is used to image the NLoS scene (occlusion not included). The synthetic apertures consist of a transceiver and a receiver that are mounted on two linear translation stages, respectively. To emulate 2D objects, a carpet and cardboard boxes are used. (b) The initial image shows that the drywall is at the correct location but the NLoS objects appear to be behind it. (c) The correct image is obtained using *mirror folding*.

The scene comprises the corner of a corridor entrance in the LoS whereas a life-sized mannequin and the inner side of the corridor (with a metallic fire extinguisher box) are in the NLoS. Using a measurement bandwidth of 270-300 GHz, Fig. 6b shows a horizontal cross-section of the obtained initial image. We observe that 1) both LoS walls are correctly imaged, including the right angle between them; 2) all the NLoS objects are revealed, including the mannequin, fire extinguisher box, and the NLoS wall; and 3) the NLoS objects are mirrored by the LoS wall. Besides, the metallic fire extinguisher box exhibits stronger backscattering, hence is brighter in the image. Fig. 6c depicts the correct image achieved by applying *mirror folding*. We notice that it properly reflects the actual measurement environment. In addition, Fig. 6d shows a vertical cross-section of the mannequin, we can easily recognize its outline and posture.

Nevertheless, to achieve high resolution and long imaging range, large apertures are needed to focus the beam to a small spot with a centimeter-scale size. Therefore, an imaging setup based on optics (e.g., lens and mirror) can be bulky with meter-scale optical components [18]. Additionally, the mechanical movement used to implement beam steering (e.g., translation and rotation) is a bottleneck of achieving a sub-second frame rate.

An alternative approach is to use antenna arrays. They can accomplish a fast-imaging rate, 3D image reconstruction, large FoV, and planar structure while maintaining a moderate aperture size to ensure good spatial resolution. To emulate the use of antenna arrays for imaging, we use a synthetic aperture radar (SAR) to evaluate THz NLoS imaging scenarios, as shown in Fig. 7a. Here, to reduce the image acquisition time, we implemented a linear SAR (2D imaging), but the approach can be extended to 3D imaging.

In this experiment, similar to the first experiment using the 3D THz radar, a drywall is used to image the NLoS scene. To fulfill 2D imaging due to the linear aperture, we use a carpet roll and cardboard boxes (inner side covered with aluminum foil to block signal penetration) as the NLoS objects. As a result, using a bandwidth of 270-300 GHz, Fig. 7b shows the reconstructed image. We can identify the drywall and two ghost objects in the shadow area. Then, using *mirror folding*, we acquire the corrected image, as depicted in Fig. 7c. We notice that it agrees with the actual setup topology well.

## Discussion

In this paper, we propose a method of THz NLoS imaging. Common building materials exhibit a very strong specular reflection in the case of THz wave illumination. This lays the foundation for THz NLoS imaging as the specular reflection can be used to deflect the signals from the imaging aperture to the invisible scene. However, it also leads to incorrect images as the NLoS objects will appear at the wrong locations. Thus, we introduced a computationally efficient post-processing method (i.e., *mirror folding*) to correct the initial images such that the hidden scene can be properly reconstructed.

To demonstrate THz NLoS imaging, we performed 2D simulations in the 290-300 GHz frequency band. Moreover, we carried out multiple measurements in the range of 270-300 GHz to further validate THz NLoS imaging. Two types of imagers are used: a 3D radar based on raster scanning and a synthetic aperture imaging system. In various scenarios, we achieved satisfying imaging results with centimeter-scale resolution.

Although this method presents for the first time the reconstruction of detailed NLoS images, further investigation is needed to improve the performance in a broad range of imaging scenarios. For example, arbitrarily shaped objects (nonmirrors) can contribute to the multiple reflection propagation and clutter the image further. Although it might be much weaker than the specular reflections of flat surfaces, it can have a deleterious effect with highly reflective materials and with short propagation distances. Additionally, if a mirror surface is semitransparent (either due to thin or low loss material), the objects appearing farther behind will not necessarily be the result of reflection but transmission. Therefore, directly applying mirror folding will result in erroneous reconstruction of the scene. This could be mitigated by generating images in both vertical and horizontal polarizations which can help distinguish the location of an object with respect to the mirror surface. Furthermore, the method can be generalized to account for curved mirror surfaces. This could be achieved by discretizing the curved surface into multiple, smaller planar segnemts before applying the proposed mirror folding method. Interestingly, the use of convex surfaces can be exploited by THz NLoS imaging to increase the FoV of the system, although with reduced spatial resolution. Future work could also investigate the efficiency of the imaging method at lower frequencies where surface roughness and spatial resolution (for a given aperture size) become smaller.

THz NLoS imaging can provide motivation and impact in other related areas. The requirement for high spatial resolution and bandwidth can achieved with advances in THz monolithic integrated circuit (TMIC) fabrication and integration techniques, planar antenna arrays, and programmable metasurfaces for electronic beamforming. Additionally, the proposed imaging method can be used to develop physics based and machine learning (ML) based radar algorithms for NLoS imaging, sensing, and detection.

## Methods

**Numerical Analysis of Multireflection Wave Propagation.** To allow the calculation of EM fields on an electrically large model, as the one encountered in a typical indoor propagation scenario, we limit the analysis on a 2D space (assume objects are infinitely uniform on the third dimension). To further simplify the model, all object surfaces are modeled as perfect electric conductors (PECs), hence material losses (other than loss due to diffuse scattering) are not included in this analysis. Besides, to numerically model the surface roughness of all the objects, we use the root mean square (RMS) height ($h$) and correlation length ($L$) to describe the roughness [27]. In this work, we assume that the walls and targets have a $h/L$ of 0.05 mm/0.2 mm and 0.1 mm/0.2mm, respectively, which are close to common building surfaces [25].

To create lines with random roughness based on the desired total length, number of points, $h$, and $L$, we first generate random data with distribution and autocovariance functions both being Gaussian, then import them into the WIPL-D 2D solver to construct a rough (wiggly) line. We note that the number of points needs to be large enough such that the segment length is smaller than $L/2$ to fulfill the Nyquist sampling theorem. Also, to achieve adequate gaussian statistics, the total length of the rough line needs to be much larger than $L$ (> 500 times larger in our model).

For this work, the THz imaging aperture consists of a 161-element linear antenna array with uniform excitation and spacing. The antennas radiate THz waves in the band of 290-300 GHz and coherently detect the backscattered signals. In addition, to avoid grating lobes in the array factor (AF), the spacing between the elements is $\lambda/2$ at the highest frequency [28], namely 0.5 mm. Thus, the total aperture size is 8 cm. Besides, all the antennas are omnidirectional and linearly polarized (normal to the topology plane). The simulations are carried out using a 2D Method-of-moments (MoM) commercial EM solver (WIPL-D 2D).

**Measurement hardware and operation.** The 3D THz radar system comprises a transceiver integrated with a diagonal horn antenna, a Teflon lens, and a motorized rotational mirror. Here, the transceiver is a vector network analyzer (VNA, model: Rohde & Schwarz ZVA24) extension module (model: Virginia Diodes WR3.4-VNAX) that measures the $S_{11}$ parameter in the WR3.4 band (220-330 GHz). The horn has a half-power beamwidth and gain of 10 degrees and 26 dB, respectively. Thus, together they emit a diverging THz beam into the free space. Besides, the customised Teflon lens has a focal length of 44.3 cm so that, when it is placed in front of the horn at a distance of 50 cm, the diverging beam is focused on a spot with a radius of approximately 1.7 cm at a distance of 3.6 m. Additionally, the customized rotational mirror is driven by two stepper motors to enable controlled rotation in both azimuth and elevation angles. Thus, the lens forms a narrow THz beam while the mirror enables beam steering (±15º).

The synthetic aperture imaging system includes a pair of THz transceiver and receiver (instead of using only one transceiver as in the 3D THz radar) for a better dynamic range. Both of them are VNA extenders in the WR3.4 band and integrated with diagonal horn antennas. The extenders are respectively mounted on two translation stages (model: Thorlabs LTS150/M) such that we translate the transceiver for 81 steps with a spacing of 0.5 mm ($\lambda/2$ at the highest frequency). Meanwhile, at each step, the receiver is translated in the same way (81 steps, 0.5 mm spacing) and coherently records the backscattered signals by measuring the $S_{21}$ parameter. In addition, the extenders are placed as close as possible to minimize the effect of aperture dislocation. As such, we emulate a linear antenna array with full multistatic measurement capability.

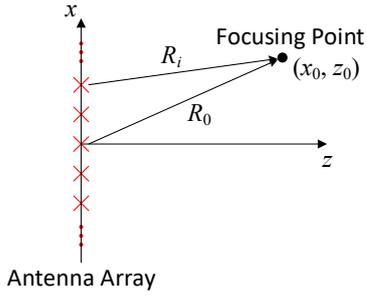

Figure 8. Illustration of the geometrical relationship between the antenna array and a focal point.

**Image reconstruction algorithm.** The 2D simulation and the synthetic aperture imaging method have the same image reconstruction algorithm. We treat the transmitting (Tx) and receiving (Rx) linear arrays as a MIMO system such that they record the backscattered signals for all the Tx-Rx combinations. To generalize the method to cases where imaging apertures sizes are comparable to the distance from the region of interest (RoI), we account for the curvature of the wavefront. Thus we use the total focusing method (TFM) [29] which exploits the information of every Tx-Rx pair in the imaging array. In contrast to ultrasound imaging, where TFM is used with pulsed signals, the examples presented in this work assume stepped frequency continuous wave (SFCW) systems. As such, the measurement results form a three-dimensional $M \times N_{Tx} \times N_{Rx}$ matrix ($\boldsymbol{S}$), where $M$ is the number of frequency points and $N_{Tx, Rx}$ are the number of Tx and Rx antennas, respectively.

As depicted in Fig. 8, to form a focused illumination at a point ($x_0$, $z_0$), we consider the distance between the focusing point and each antenna element independently. Thereby, the relative phase delay of the signal at the $i$-th antenna on the array is:

$$\varphi_i = \frac{2\pi}{\lambda}(R_i - R_0) \quad (1)$$

TABLE I
Algorithm and Pseudocode for Image Reconstruction Using Mirror Folding

1. *Raw image acquisition*: The THz camera generates a 3D matrix forming the initial image (cross-range and range). The numerical value assigned to each voxel corresponds to the intensity of the backscattered signals. Additionally, to compensate for the free space path loss, each voxel's intensity is normalized with the $r^2$ factor ($r$ is distance of voxel from the THz camera's reference point). Besides, to extract a 2D image (horizontal cross-section) as we demonstrated in this work, we only keep the pixel with the maximum value in every vertical cross-range column.
2. *Determine the mirror pixels*: Set a threshold $T$ such that only the pixels with an intensity $>T$ are considered as objects whereas others are recognized as free space. Then, apply the Hough transform on the objects to determine the linear segments that correspond to flat surfaces (mirrors) [30]. The remaining pixels above threshold $T$ are assumed to be nonmirror (objects) and do not contribute to multipath propagation.
3. *Hierachical mirror classification based on the number of bounces*: As shown in the diagram of Fig. 9, the detected flat surfaces in the LoS are designated as the "parent" mirrors (e.g., *image.mirror().line()*). Accordingly, other identified flat surfaces at longer range distances (i.e., appear to be behind LoS mirrors) become "child" mirrors. For example, for a second bounce mirror the corresponding pixel line will be
    *image.mirror(i).mirror(k).line()*
    where $i$ corresponds to the i[th] LoS mirror (first generation) and $k$ corresponds to the k[th] next generation mirror of the i[th] LoS mirror. This way, the total number of bounces determines the number of mirror generations.
4. *Hierachical (nonmirror) object classification:* After the mirror detection and classification, the remaining pixels above the threshold $T$ are classified hierarchically similar to the mirror classification process (step 3), as shown in diagram of Fig. 9. As such, each mirror determines a shadow region with child mirrors and/or nonmirror objects. For example, *image.mirror(1).mirror(3).objects()* stores the indices of the pixels in the shadow area of the 3[rd] child mirror behind the 1[st] LoS mirror.
5. *Apply mirror folding on every branch of the tree structure:*
    $j$ = maximum number of bounces
    while $j > 0$
       for $t <$ number of mirrors of j[th] bounce
          (i) Rotate all objects around the line defined by the parent mirror (find the new coordinates)
          (ii) Assign the new pixel coordinates to the previous generation. For example, the updated coordinates of *image.mirror(2).mirror(1).objects()* will be stored into the *image.mirror(2).objects()* method.
       $t \leftarrow t+1$
    $j \leftarrow j-1$
6. Finally, *image.objects()* contains the pixels of the reconstucted scene.

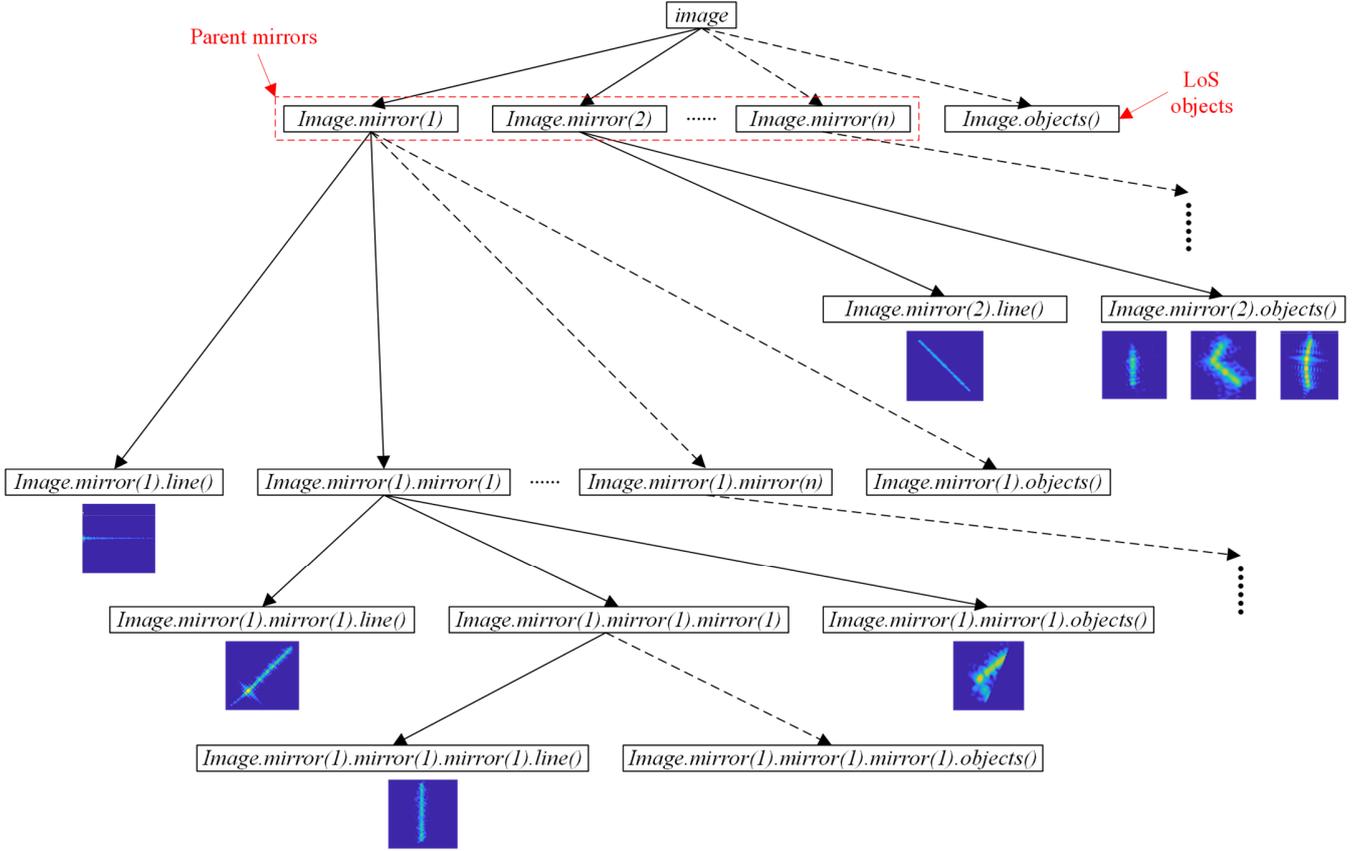

Figure 9. Tree structure of the hierarchical classification of mirrors and other objects on the initial THz image (see Fig. 4a).

with

$$R_0 = \sqrt{x_0^2 + z_0^2} \quad (2)$$

$$R_i = \sqrt{(x_i - x_0)^2 + z_0^2} \quad (3)$$

where λ is the wavelength, $R_0$ and $R_i$ are the distances of the focusing point from the array center (reference element, zero phase) and the $i$-th element, respectively. Thus, based on $M$ frequency points, we accquire two phase matrices $\boldsymbol{\Phi}_{Tx}$ and $\boldsymbol{\Phi}_{Rx}$, with a size of $M \times N_{Tx}$ and $M \times N_{Rx}$, for the Tx and Rx arrays, respectively. The columns of the matrices represent the phase delays of the corresponding elements at every frequency. Then, beam focusing by the Tx and Rx arrays is achieved by applying the necessary phase delays:

$$\boldsymbol{S}_0 = \sum_{\text{Dim}_2}\left(\sum_{\text{Dim}_2} \boldsymbol{S}_k \circ \boldsymbol{\Phi}_{Tx}\right) \circ \boldsymbol{\Phi}_{Rx} \quad (4)$$

Here, the operator ∘ stands for the Hadamard product and $\boldsymbol{S}_k$ is the $k$-th 2D matrix along the third dimension of $\boldsymbol{S}$ ($k$=1, 2, ..., $N_{Rx}$). Also, Dim$_2$ denotes that the summation is operated in the second dimension of the matrix. As a result, we obtain a frequency domain vector $\boldsymbol{S}_0$ with $M$ elements. By applying the inverse Fourier transform (IFT), we acquire the time domain signal:

$$\boldsymbol{S}_t = \text{IFT}[\boldsymbol{S}_0 \circ \boldsymbol{a}] \quad (5)$$

To achieve lower side lobes in the range direction, we taper $\boldsymbol{S}_0$ before calculating the IFT. For simplicity and without loss of generality, we use the triangular function $\boldsymbol{a}$, as determined by:

$$a[n] = 1 - \frac{2}{M-1}\left|n - \frac{M+1}{2}\right| \quad (6)$$

The time domain vector $\boldsymbol{S}_t$ is then converted to distance (range) based on the wave velocity in free space ($c_0$). According to the sampling theorem, the maximum unambiguous detectable range is $c_0/(2\Delta f)$, where $\Delta f$ is the frequency step. However, since $\boldsymbol{S}_t$ is calculated for a single focusing point, we only keep the value that corresponds to the range of this point in $\boldsymbol{S}_t$. Finally, following the same process, we calculate the backscattered signal from every focusing point ($x, z$) in the RoI and combine them to form the reconstructed image.

On the other hand, the 3D THz radar has a simpler image reconstruction algorithm. At each scanning angle, the transceiver measures the $S_{11}$ parameter in the frequency-domain. Then we convert it to time-domain using the IFT, thus accquiring the range information based on $c_0$. As a result, the radar system produces a 3D matrix with a size of $N_a \times N_e \times M$, where $N_a$ and $N_e$ are the number of scanning angles in the azimuth and elevation planes, respectively, and $M$ is the number of frequency points (i.e., discretized points along the range direction). Therefore, each matrix element corrsponds to a voxel and a 3D image is obtained by plotting the magnitude of all the voxels.

**Mirror folding algorithm.** The initial image captured by the imaging aperture (e.g., THz camera) is processed to reveal the correct geometry and pose of NLoS objects and scenes. We are using an inverse approach to the mirror unfolding approach used in ray tracing methods in computer graphics. We term this approach *mirror folding* and it is summarized in table I. After the initial THz image is acquired, the method detects linear segments of minimum length and classifies them as mirrors. The relative position of every mirror determines the number of bounces that the signal has encountered by assuming that surfaces are opaque to the THz waves. This multibounce propagation is modeled as a hierachical tree (see Fig. 9) where behind every mirror the subsequent pixels belong to the respective mirrors generation tree. The number of generations depends on the number of bounces. Finally, by systematically applying mirror folding, we reconstruct the correct geoemetry and pose of both LoS and NLoS scenes. Although for this work the reconstruction has been limited to 2D images and flat surfaces, it can be easilty generalized for 3D images and curved surfaces.

**References**

1. M. Dehmollaian and K. Sarabandi, "Refocusing Through Building Walls Using Synthetic Aperture Radar," in *IEEE Transactions on Geoscience and Remote Sensing*, vol. 46, no. 6, pp. 1589-1599, June 2008, doi: 10.1109/TGRS.2008.916212.
2. K. Chetty, G. E. Smith and K. Woodbridge, "Through-the-Wall Sensing of Personnel Using Passive Bistatic WiFi Radar at Standoff Distances," in *IEEE Transactions on Geoscience and Remote Sensing*, vol. 50, no. 4, pp. 1218-1226, April 2012, doi: 10.1109/TGRS.2011.2164411.
3. T. Li, L. Fan, M. Zhao, Y. Liu and D. Katabi, "Making the Invisible Visible: Action Recognition Through Walls and Occlusions," *2019 IEEE/CVF International Conference on Computer Vision (ICCV)*, 2019, pp. 872-881, doi: 10.1109/ICCV.2019.00096.
4. F. C. S. da Silva, A. B. Kos, G. E. Antonucci, J. B. Coder, C. W. Nelson, and A. Hati, "Continuous-capture microwave imaging," *Nat Commun*, vol. 12, no. 1, p. 3981, Dec. 2021, doi: 10.1038/s41467-021-24219-0.
5. N. Scheiner et al., "Seeing Around Street Corners: Non-Line-of-Sight Detection and Tracking In-the-Wild Using Doppler Radar," in *2020 IEEE/CVF Conference on Computer Vision and Pattern Recognition (CVPR)*, Seattle, WA, USA, Jun. 2020, pp. 2065–2074. doi: 10.1109/CVPR42600.2020.00214.
6. A. Sume, M. Gustafsson, M. Herberthson, A. Jannis, S. Nilsson, and J. Rahm, "Radar Detection of Moving Targets Behind Corners," in *IEEE Transactions on Geoscience and Remote Sensing*, vol. 49, no. 6, pp. 2259-2267, June 2011.
7. A. Velten, T. Willwacher, O. Gupta, A. Veeraraghavan, M. G. Bawendi, and R. Raskar, "Recovering three-dimensional shape around a corner using ultrafast time-of-flight imaging," *Nat Commun*, vol. 3, no. 1, p. 745, Jan. 2012, doi: 10.1038/ncomms1747.
8. G. Gariepy, *et al.*, "Detection and tracking of moving objects hidden from view," *Nature Photonics*, 10, 23–26, 2016, doi: 10.1038/nphoton.2015.234.
9. M. O'Toole, D. B. Lindell, and G. Wetzstein, "Confocal non-line-of-sight imaging based on the light-cone transform," *Nature*, vol. 555, no. 7696, pp. 338–341, Mar. 2018, doi: 10.1038/nature25489.
10. X. Liu *et al.*, "Non-line-of-sight imaging using phasor-field virtual wave optics," *Nature*, vol. 572, no. 7771, pp. 620–623, Aug. 2019, doi: 10.1038/s41586-019-1461-3.
11. X. Liu, J. Wang, Z. Li, Z. Shi, X. Fu, and L. Qiu, "Non-line-of-sight reconstruction with signal–object collaborative regularization," *Light Sci Appl*, vol. 10, no. 1, p. 198, Dec. 2021, doi: 10.1038/s41377-021-00633-3.
12. J. H. Nam *et al.*, "Low-latency time-of-flight non-line-of-sight imaging at 5 frames per second," *Nat Commun*, vol. 12, no. 1, p. 6526, Dec. 2021, doi: 10.1038/s41467-021-26721-x.
13. C. Thrampoulidis *et al.*, "Exploiting Occlusion in Non-Line-of-Sight Active Imaging," *IEEE Transactions on Computational Imaging*, vol. 4, no. 3, pp. 419–431, Sep. 2018, doi: 10.1109/TCI.2018.2829599.
14. Y. Cui and G. C. Trichopoulos, "Comparison of Propagation Losses in THz and Optical Non-Line-of-Sight Imaging," *2019 IEEE International Symposium on Antennas and Propagation and USNC-URSI Radio Science Meeting*, Atlanta, GA, USA, 2019, pp. 1473-1474, doi: 10.1109/APUSNCURSINRSM.2019.8888705.
15. C. Jansen *et al.*, "Diffuse Scattering From Rough Surfaces in THz Communication Channels," in *IEEE Transactions on Terahertz Science and Technology*, vol. 1, no. 2, pp. 462-472, Nov. 2011, doi: 10.1109/TTHZ.2011.2153610.
16. S. k. Doddalla and G. C. Trichopoulos, "Non-Line of Sight Terahertz Imaging from a Single Viewpoint," *2018 IEEE/MTT-S International Microwave Symposium - IMS*, Philadelphia, PA, 2018, pp. 1527-1529.
17. G. C. Trichopoulos, H. L. Mosbacker, D. Burdette and K. Sertel, "A Broadband Focal Plane Array Camera for Real-time THz Imaging Applications," in *IEEE Transactions on Antennas and Propagation*, vol. 61, no. 4, pp. 1733-1740, April 2013, doi: 10.1109/TAP.2013.2242829.
18. K. B. Cooper, R. J. Dengler, N. Llombart, B. Thomas, G. Chattopadhyay and P. H. Siegel, "THz Imaging Radar for Standoff Personnel Screening," in *IEEE Transactions on Terahertz Science and Technology*, vol. 1, no. 1, pp. 169-182, Sept. 2011, doi: 10.1109/TTHZ.2011.2159556.
19. D. M. Sheen, D. L. McMakin and T. E. Hall, "Three-dimensional millimeter-wave imaging for concealed weapon detection," in *IEEE Transactions on Microwave*

*Theory and Techniques*, vol. 49, no. 9, pp. 1581-1592, Sept. 2001, doi: 10.1109/22.942570.
20. M. Aladsani, A. Alkhateeb and G. C. Trichopoulos, "Leveraging mmWave Imaging and Communications for Simultaneous Localization and Mapping," *ICASSP 2019 - 2019 IEEE International Conference on Acoustics, Speech and Signal Processing (ICASSP)*, Brighton, United Kingdom, 2019, pp. 4539-4543, doi: 10.1109/ICASSP.2019.8682741.
21. B. G. Kashyap, P. C. Theofanopoulos, Y. Cui and G. C. Trichopoulos, "Mitigating Quantization Lobes in mmWave Low-Bit Reconfigurable Reflective Surfaces," in *IEEE Open Journal of Antennas and Propagation*, vol. 1, pp. 604-614, 2020, doi: 10.1109/OJAP.2020.3034049.
22. R. G. Baraniuk, "Compressive Sensing [Lecture Notes]," in *IEEE Signal Processing Magazine*, vol. 24, no. 4, pp. 118-121, July 2007, doi: 10.1109/MSP.2007.4286571.
23. M. Grzegorzek, C. Theobalt, R. Koch, and A. Kolb, Time-of-Flight and Depth Imaging: Sensors, Algorithms and Applications. 1st edition. Berlin, Heidelberg, Germany: Springer, 2013.
24. O. Gupta, T. Willwacher, A. Velten, A. Veeraraghavan, and Ramesh Raskar, "Reconstruction of hidden 3D shapes using diffuse reflections," *Opt. Express* 20, 19096-1910, 2012.
25. Y. Cui, M. K. Immadisetty and G. C. Trichopoulos, "Evaluating the Properties of Millimeter- and THz Wave Scattering from Common Rough Surfaces," *2020 IEEE International Symposium on Antennas and Propagation and North American Radio Science Meeting*, 2020, pp. 1117-1118, doi: 10.1109/IEEECONF35879.2020.9329367.
26. M. Khatun, C. Guo, D. Matolak and H. Mehrpouyan, "Indoor and Outdoor Penetration Loss Measurements at 73 and 81 GHz," 2019 *IEEE Global Communications Conference (GLOBECOM)*, 2019, pp. 1-5, doi: 10.1109/GLOBECOM38437.2019.9013945.
27. P. Beckmann and A. Spizzichino, The Scattering of Electromagnetic Waves from Rough Surfaces. Norwood, MA, USA: Artech House, 1987.
28. C. A. Balanis, Antenna Theory: Analysis and Design, 4th edition. Hoboken, NJ, USA: John Wiley & Sons, 2016.
29. C. Holmes, B. Drinkwater, and P. Wilcox, "The post-processing of ultrasonic array data using the total focusing method," *Insight - Non-Destructive Testing and Condition Monitoring*, vol. 46, no. 11, pp. 677-680(4), Nov. 2004, doi: 10.1784/insi.46.11.677.52285.
30. R. O. Duda and P. E. Hart, "Use of the Hough transformation to detect lines and curves in pictures," *Communications of the ACM*, vol. 15, no. 1, pp. 11-15, Jan. 1972.